\documentclass[pra, aps, amsmath, amssymb, twocolumn, superscriptaddress, eqsecnum, nofootinbib]{revtex4-2}

\usepackage{amsmath}
\usepackage{graphicx}
\usepackage{mathrsfs}
\usepackage{color}
\usepackage{hyperref}

\hypersetup{
	colorlinks = true,
	urlcolor   = blue,
	linkcolor  = red,
	citecolor  = blue
}

\begin{document}

\title{Single collective excitation of an atomic array trapped along a waveguide: \\a study of cooperative emission for different atomic chain configurations}

\author{V.A. Pivovarov}
\affiliation{Physics Department, St.-Petersburg Academic University, Khlopina 8, 194021 St.-Petersburg, Russia}
\affiliation{Center for Advanced Studies, Peter the Great St-Petersburg Polytechnic University, 195251, St.-Petersburg, Russia}
\author{L.V. Gerasimov}
\affiliation{Quantum Technologies Center, M.V.~Lomonosov Moscow State University, Leninskiye Gory 1-35, 119991, Moscow, Russia}
\affiliation{Center for Advanced Studies, Peter the Great St-Petersburg Polytechnic University, 195251, St.-Petersburg, Russia}
\author{J. Berroir}
\affiliation{Laboratoire Kastler Brossel, Sorbonne Universit\'e, CNRS, ENS-Universit\'e PSL, Coll\`{e}ge de France, 4 place Jussieu, 75005 Paris, France}
\author{T. Ray}
\affiliation{Laboratoire Kastler Brossel, Sorbonne Universit\'e, CNRS, ENS-Universit\'e PSL, Coll\`{e}ge de France, 4 place Jussieu, 75005 Paris, France}
\author{J. Laurat}
\affiliation{Laboratoire Kastler Brossel, Sorbonne Universit\'e, CNRS, ENS-Universit\'e PSL, Coll\`{e}ge de France, 4 place Jussieu, 75005 Paris, France}
\author{A. Urvoy}\email{alban.urvoy@sorbonne-universite.fr}
\affiliation{Laboratoire Kastler Brossel, Sorbonne Universit\'e, CNRS, ENS-Universit\'e PSL, Coll\`{e}ge de France, 4 place Jussieu, 75005 Paris, France}
\author{D.V. Kupriyanov}\email{kupriyanov@quantum.msu.ru}
\affiliation{Quantum Technologies Center, M.V.~Lomonosov Moscow State University, Leninskiye Gory 1-35, 119991, Moscow, Russia}
\affiliation{Center for Advanced Studies, Peter the Great St-Petersburg Polytechnic University, 195251, St.-Petersburg, Russia}

\date{\today}
\begin{abstract}
\noindent Ordered atomic arrays trapped in the vicinity of nanoscale waveguides offer original light-matter interfaces, with applications to quantum information and quantum non-linear optics. Here, we study the decay dynamics of a single collective atomic excitation coupled to a waveguide in different configurations. The atoms are arranged as a linear array and only a segment of them is excited to a superradiant mode and emits light into the waveguide. Additional atomic chains placed on one or both sides play a passive role, either reflecting or absorbing this emission. We show that when varying the geometry, such a one-dimensional atomic system could be able to redirect the emitted light, to directionally reduce or enhance it, and in some cases to localize it in a cavity formed by the atomic mirrors bounding the system.
\end{abstract}

\maketitle

\section{Introduction}

\noindent Developing and harnessing hybrid platforms where neutral atoms interact with guided photons is presently a very active area of research, opening new opportunities for light-matter interactions \cite{RMPKimble}. This emerging field, known as waveguide quantum electrodynamics, has led to a variety of theoretical proposals and to experimental demonstrations with nanofiber-trapped atoms \cite{Nieddu2016,Solano2017}, ranging from the realization of optical memories \cite{memoryLKB,memoryVienna} to the observation of superadiance and collective non-linear effects \cite{Solano,Vienna}. Recently, the generation of a waveguide-coupled atomic excitation that can be read out as a guided single photon was also demonstrated \cite{NatureLKB}. Novel platforms are actively studied, including with one-dimensional photonic-crystal waveguides that would offer the unique possibility of exploring bandgap physics \cite{Caltech1D,ChangPRX,Beguin}.

In such trapped atomic arrays, when the guided mode interacts with a periodically ordered chain of atoms fulfilling the Bragg resonance condition, the ensemble of atoms can behave as a Bragg mirror~\cite{Deutsch1995}. Such Bragg reflection was observed with several thousands nanofiber-trapped atoms \cite{CGCGSKL2016,SBKISMPA2016}. This system can be used to explore quantum non-linear optics and the realization of a single-photon gate \cite{Albrecht2017}. Cooperative effects in this configuration have also raised a large interest, with not only extremely subradiant states \cite{Sheremet} but also selectively radiant states that may enable exponential improvement in photon storage fidelities \cite{Asenjo}.

Given theses capabilities, a natural extension is to combine two atomic Bragg mirrors with additional atoms placed inbetween to form an atomic resonator. This configuration was theoretically explored in \cite{Chang2012}, with a single atom strategically placed inside a very short atomic resonator at one of its anti-nodes. A vacuum Rabi splitting arising from the enhancement of the density of states was predicted, hence demonstrating that strong coupling can be achieved, akin to cavity quantum electrodynamics (QED) with classical mirrors. The transition from a short atomic resonator, i.e., in the regime where propagation delay is negligible with respect to the mirror response time, to long ones was detailed in \cite{Scarani2016}, showing for long resonators a complete analogy to conventional cavity-QED with enhanced photon lifetime.

In this paper, as sketched in Fig.~\ref{fig1}, we now consider an ensemble of atoms placed inside the resonator and initially prepared in a superradiant atomic mode that cooperatively emits into the waveguide. As in~\cite{Scarani2016}, we consider both the Markovian and non-Markovian regime, confirming the key differences between the two. We follow the conventionally used definitions where the non-Markovian regime is associated with non-negligible retardation in the interactions through virtual photon exchange with respect to the system's dynamics. We introduce novel aspects, such as the effect of disorder in the atomic mirrors. We also examine novel configurations, such as one-sided case where one atomic mirror is removed. Another explored configuration is a system in which the superradiant mode is prepared in a segment of a larger atomic Bragg mirror, resulting in two timescales governing the dynamics. Our treatment relies on a microscopic ab-initio theory in the presence of a nanostructure and intrinsically includes spontaneous emission and dipole-dipole coupling between the atoms in free-space.

The paper is organized as follows. In Sec. \ref{sec2}, we first review the system under study and the general theoretical framework. Section \ref{sec3} and \ref{sec4} provide theoretical simulations of the decay dynamics, for different configurations. Section \ref{sec3} considers the Markovian regime with short distances between atomic Bragg mirrors while Sec. \ref{sec4} focuses on the non-Markovian one, which maps very well to the conventional cavity-QED setting even for an excitation stored in an ensemble of atoms.

\section{Fundamentals: system and model} \label{sec2}
\noindent In this section, we first present the considered scheme for coupling an atomic array to a nanoscale dielectric waveguide and we introduce the parameters that are experimentally relevant.
Then we focus on the theoretical model describing the atomic excitation temporal evolution and the photonic modes emerging from this evolution. We conclude this section by presenting the decay dynamics for a chain of atoms sharing an initial single collective excitation, the base case where left and right atomic Bragg mirrors are absent.

\subsection{An example: a nanofiber-trapped atomic array}

\begin{figure}[tp]
\includegraphics[width=8.6cm]{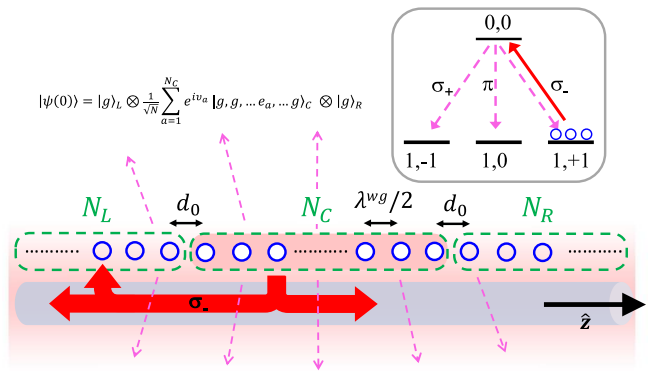}
\caption{Schematic representation of the system.An atomic array trapped near a sub-wavelength dielectric waveguide is split in three segments (left, center, right) separated by a distance $d_0$. Atoms within each segment are spaced by $\lambda^{\mathrm{wg}}/2$ (unless specified otherwise). Disorder will also be considered for the left and right atomic mirrors. All atoms have a tripod level scheme, as shown in the inset, and are initially spin polarized in the $\mathbf{\hat{z}}$ direction, albeit for the central segment that contains one excitation with the proper phase factors for cooperative emission into the waveguide. The atoms can emit light in free space and in the waveguide in all three polarizations (dashed pink arrows). The red arrows specifically highlight the coupling of an arbitrary atomic pair on $\sigma_{-}$ atomic transition via the guided field, which is the only one that can cooperatively couple to the spin-polarized ground state atoms via the Rayleigh channel and thus undergo Bragg reflection from the segments. The aim of this paper is to study the dynamics of the decay process in various configurations.
}
\label{fig1}%
\end{figure}%

\noindent In our study, the atoms are modeled by a tripod energy configuration, with the following spin angular momenta and projections: $F_{0} = 1$, $M_{0} = 0$, $\pm1$ (ground state $|m\rangle$) and $F, M = 0, 0$ (excited state $|n\rangle$). Such an energy configuration can be found in the hyperfine manifold of $^{87}$Rb, but more importantly, this scheme allows us to model a rather realistic example of an open system in which an emitted photon can experience either Rayleigh or Raman scattering. As we will show, even with such an open system, the cooperative processes indeed mainly develop via the Rayleigh scattering channel, whose main role is to transport a polaritonic excitation along the waveguide.

The quantization axis is taken along the axis of the waveguide, following the approach of Ref.~\cite{Pivovarov2020}, which is a practical configuration for a tripod scheme, and offers the benefits of a cylindrical symmetry. The atoms are all prepared in the $|F_{0},+1\rangle$ state (spin oriented along the waveguide).\footnote{This choice is different than in most experimental implementations, where the quantization axis is taken to be normal to the waveguide and typically in the plane of the atoms and waveguide. If instead of circular we used linear polarizations along either $x$ or $y$ axes, and superposed our atoms between $m=\pm 1$ states, we would obtain similar results to those discussed below.} Subsequently, a weak and short coherent external light pulse excites the active atoms and creates a single collective excitation in a superradiant mode in the atomic array.

As a typical experimental example, in this study we consider the $|F_{0}=0\rangle \rightarrow |F=1\rangle$ transition of the $^{87}$Rb D2 line, with a vacuum transition wavelength $\lambda_{0}~=~780$~nm, coupled to a silica nanofiber of radius $a \sim 200$ nm. The atoms are separated from the surface by half the fiber radius $a/2\sim 100$~nm. This configuration leads to a single-atom coupling ratio of $\beta = \gamma_{\textrm{1D}}/\gamma\sim 0.1$, where $\gamma_{\textrm{1D}}$ and $\gamma$ are the radiative decay rates into the guided mode and in free space (undisturbed), respectively \cite{CGCGSKL2016,LeKien2014,PLGPCLK2018}.

\subsection{Microscopic description of the excitation decay}
\label{sec2B}
\noindent We first consider an atomic chain, with a single collective excitation shared among all $N_C$ central atoms at a given initial time $t=0$ (central atomic chain in Fig. \ref{fig1}).
In the subsequent quantum description, such a collective atomic state, which we denote as $|\psi(0)\rangle$, should be expanded in the complete basis set by including the vacuum environment of the field modes.
Since this state is not an eigenstate of the Hamiltonian of the system, it further experiences a joint dynamics with the field modes, such that up to arbitrary time $t$ it evolves to

\begin{equation}
|\psi(t)\rangle=\exp\left(-\frac{i}{\hbar}\hat{H}\,t\right)|\psi(0)\rangle\equiv U(t,0)\,|\psi(0)\rangle%
\label{2.1}%
\end{equation}
where $|\psi(t)\rangle$ is a polariton-type entangled state that superposes a single atomic excitation and a single-photon occupation of the field modes. The decay of the excitation can be traced by the time-dependent probability $p=p(t)$ given by
\begin{equation}
p(t)=\mathrm{Tr}\hat{P}\,\hat{\rho}(t) \quad \textrm{with} \quad \hat{\rho}(t)=|\psi(t)\rangle\langle\psi(t)|.
\label{2.2}%
\end{equation}
The operator $\hat{P}$ defined as
\begin{eqnarray}
\lefteqn{\hspace{-0.8cm}\hat{P}=\sum_{a=1}^{N}\;\sum_{\{m_j\},j\neq a}\;\sum_{n}%
|m_1,\ldots,m_{a-1},n,m_{a+1},\ldots m_N\rangle}%
\nonumber\\%
&&\hspace{-0.5cm}\langle m_1,\ldots,m_{a-1},n,m_{a+1},\ldots m_N|\times|0\rangle\langle 0|_{\mathrm{Field}},%
\label{2.3}%
\end{eqnarray}
projects the system density matrix $\hat{\rho}(t)$ onto the Hilbert subspace of the atomic excitation, where the photonic field is vacuum and the atomic state is any superposition of singly excited states with one specific $a$'th atom (with $a$ running from $1$ to $N$) occupying a Zeeman sublevel $|n\rangle$ of the excited state, and all $N-1$ other atoms occupying a Zeeman sublevel $|m_j\rangle$ ($j \neq a$) of the ground state. Note that the initial state is one of such states. The probability to obtain the system in its initial state can be found by projecting the final state $\hat{\rho}(t)$ onto its initial state with $\hat{P}\to\hat{P}_0=|\psi(0)\rangle\langle\psi(0)|$ in (\ref{2.2}). In this case, we denote the observation probability as $p_0(t)$.

As $\hat{P}=\hat{P}^2$ and $\hat{P}|\psi(0)\rangle=|\psi(0)\rangle$, the excitation probability can be expressed in the following form
\begin{equation}
p(t)=\mathrm{Tr}\,\hat{P}\,U(t,0)\hat{P}\,\hat{\rho}(0)\,\hat{P}\,U^{\dagger}(t,0)\hat{P}.%
\label{2.4}%
\end{equation}
This basic equation describes the microscopic dynamics of the decay process in a multiatomic system, where the optical excitation can be virtually re-emitted many times. We considered here only the transformation of pure states, but the derivation can be straightforwardly generalized towards transformation of an arbitrary initial mixed state $\rho(0)$. The evolution operator $U(t,0)$ can be found in previous work \cite{SMLK2012,KSH2017,SKGLK2015,PLGPCLK2018,Pivovarov2020} and is summarized in Appendix \ref{Appendix_Dynamics}.

Let us point out that the projection of the evolutionary operator on a single excitation subspace physically means that no doubly excited states play a role in the propagation dynamics through the waveguide. However including doubly excited states is necessary in our model to account for the near field virtual interaction giving rise to a static dipole-dipole coupling for any atom with its proximal neighbors.

\subsection{Light emitted into the waveguide}

\noindent After describing the general theoretical framework that gave us access to the evolution of the atomic populations, let us now extend it to also extract the temporal profile of the light emitted into the waveguide.

From the point of view of scattering theory we deal here with an example of a process, where the wave-packet of an initial state of the compound system evolves into the final state of two uncoupled subsystems belonging to a particular scattering channel. In accordance with general principles of the scattering theory, the dynamics of such an incomplete scattering process can be described with the formalism of the M{\o}ller  operators: the system's states are transformed either from infinite past (operator $\hat{\Omega}^{(+)}$) or from infinite future (operator $\hat{\Omega}^{(-)}$) to a compound state considered at the conventional initial moment $t=0$ \cite{GoldbergerWatson64}. In the interaction representation, the corresponding asymptotic transformation from the initial state $|\psi(0)\rangle$ (excited atoms) to the outgoing state $|\psi\rangle_{\mathrm{out}}$ (emitted photon), is given by
\begin{eqnarray}
|\psi\rangle_{\mathrm{out}}&=&\hat{\Omega}^{(-)\dagger}|\psi(0)\rangle%
\nonumber\\%
\hat{\Omega}^{(\pm)}&=&\lim_{\tau\to\infty}\exp\left(\mp\frac{i}{2\hbar}\hat{H}\,\tau\right)\exp\left(\pm\frac{i}{2\hbar}\hat{H}_0\tau\right)%
\nonumber\\%
&=& \lim_{\tau\to\infty} U(\pm\tau/2,0)U_0(\mp\tau/2,0)%
\label{2.5}
\end{eqnarray}
where $H$ is the system Hamiltonian and $H_0$ is the channel Hamiltonian corresponding to the separated subsystems. In our case the output channel consists of the emitted photon and the de-excited atomic chain. Since the latter exists in its ground state, then in the rotating wave approximation (RWA), we can ignore in $H_0$ any coupling with the quantized field.

The M{\o}ller  operators are nontrivial mathematical objects which map the states belonging to the continuous spectrum of the compound system onto the separated states for the particular scattering channel. Although the operators are constructed as a product of two unitary operators $U(\ldots)$ and $U_0(\ldots)$ describing the interacting and non-interacting dynamics respectively, if the compound system had a set of stable states belonging to its discrete spectrum, the product considered at infinite $\tau\to\infty$ generally would not tend to a unitary operator. In our system there are no stable and localized coupled excited states of the field and the atoms and the M{\o}ller operators are unitary. We detail the operator $\hat{\Omega}^{(-)\dagger}$, contributing to the transformation (\ref{2.5}), in Appendix \ref{Appendix_Moller}.

The asymptotic state $|\psi\rangle_{\mathrm{out}}$ can be decomposed into the infinite set of the states $|f\rangle\equiv|G,s\rangle$. Here $|G\rangle=|m_1,\ldots,m_N\rangle$ represents any of the collective atomic ground eigenstates, and the quantum number $s$ specifies the freely-propagating field modes modified by the presence of the waveguide. One particular mode is the guided mode $s=\sigma,k$ parameterized by an azimuthal number $\sigma=\pm 1$ and a longitudinal wavenumber $k$. Some details regarding the description of the field and of the atom-field interaction Hamiltonian near a nanostructure are discussed in Appendix \ref{Appendix_Interaction}. The transition amplitude to a particular $f$-th final state is eventually given by
\begin{eqnarray}
M_{f0}&=&\langle f|\hat{\Omega}^{(-)\dagger}|\psi(0)\rangle%
\nonumber\\%
&=&i\sqrt{2\pi\hbar\omega_{s}}\sum_{a=1}^{N}\;\sum_{n}\left(\mathbf{d}\!\cdot\!\mathbf{D}^{(s)}(\mathbf{r}_a)\right)_{nm_a}^{*}%
\nonumber\\%
&\times&\langle\ldots m_{a-1},n,m_{a+1}\ldots |\tilde{\hat{R}}(E_f+i0)|\psi(0)\rangle,%
\nonumber\\
&&\label{2.6}%
\end{eqnarray}
where $\omega_s$ is the frequency of the photon outgoing to the $s$-th mode. The mode, associated with the photon injected into the waveguide, contributes through the spatial profile of its displacement field $\mathbf{D}^{(s)}(\mathbf{r})$ subsequently taken at the position $\mathbf{r}=\mathbf{r}_a$ of each atom in the chain.

The spatial profile $\alpha=\alpha(z)$ for a pulse outgoing the system within any azimuthal guided mode, and when the atoms occupy any ground state, can be constructed as
\begin{equation}
|\alpha(z)|^2=\sum_{f,f'}M_{f0}M^{\ast}_{f'0}\,\mathrm{e}^{i(k-k')z}%
\label{2.7}
\end{equation}
where $f=f'$ excepting the longitudinal wavenumbers $k$ and $k'$. In accordance with the general concept of the scattering theory, the transformation (\ref{2.5}) concerns only spatial profile of the asymptotic wave emerging the system. Physically we can associate it with free dynamics of the outgoing light wavepacket by substituting $z\to z-v_g\tau$, where $v_g$ is the group velocity.\footnote{Even with negligible dispersion the group velocity is an important parameter of the theory, see \cite{PLGPCLK2018,Pivovarov2020}, and typically it is about $v_g \sim 0.7\, c$ for a nanofiber.}

\subsection{The base case: decay dynamics for a chain with a single atomic excitation}

\noindent We consider here the case of a single segment of $N_C$ atoms, with the initial atomic state given by
\begin{equation}
|\psi(0)\rangle=\frac{1}{\sqrt{N_C}}\sum_{a=1}^{N_C}\mathrm{e}^{i\vartheta_{a}}|g,g,\ldots,e_{a},\ldots,g\rangle.
\label{2.8}
\end{equation}
For the sake of simplicity, $|g\rangle\equiv|F_0,+1\rangle$ denotes the atomic ground state and $|e\rangle\equiv|F,0\rangle$ the excited state. The ket-vector in the right-hand side of (\ref{2.8}) can incorporate the occupied atomic states from all the segments -- central, left and right -- but the excitation is shared only by the atoms of the central segment, as clarified in Fig.~\ref{fig1}. In order for the atoms' emission to constructively interfere into the waveguide, each phase shift $\vartheta_a$ must be matching the retardation phase inside the waveguide at the position of the $a$-th atom. If the atoms are separated by a multiple of a half-wavelength of the guided mode $\lambda^{\mathrm{wg}}/2$, the phases take the values $\vartheta_a=\pi$ (odd multiplier) and $\vartheta_a=0$ (even multiplier) and the radiation is symmetrically emitted in both the waveguide directions. $|\psi(0)\rangle$ is thus a Dicke state and the superradiant mode of the central segment, emitting cooperatively into the waveguide. Experimentally, the state (\ref{2.8}) can be prepared by a short external pulse excitation additionally phase modulated along the waveguide, or through heralding a spin excitation~\cite{NatureLKB}.

For a separation between the atoms given by a multiple of a half-wavelength of the guided mode -- the well-known atomic Bragg mirror case -- the system is open from both sides and exhibits a cooperative exponential decay, with a rate given by $\Gamma_C\simeq N_C\,\gamma_{\textrm{1D}}/2$. The scaling rate $\gamma_{\textrm{1D}}$ coincides with the evaluation of the single particle self-energy part in \cite{PLGPCLK2018}. Let us note that the $1/2$ factor results here from the polarization sensitivity of the cooperative emission on the $\sigma_-$ transition, see Fig.~\ref{fig1}.\footnote{Here one should not confuse the $\sigma_{-}$ transition with the azimuthal mode $\sigma=-1$. In the emission process the atoms are coupled with both the modes $\sigma=\pm 1$ and each of the modes overlaps all three transitions $\sigma_{-},\pi,\sigma_{+}$. Nevertheless the emission of a single atom into the waveguide via the $\pi$-transition is typically much weaker than via the $\sigma_{\pm}$-transitions that justifies the $1/2$ factor, see \cite{PLGPCLK2018} ( Eq.~(3.5)) and related comments there.}

We investigate how the emission by such a chain would change in different conditions. Specifically, this base case of a single excited chain will be compared with two other geometries, where we extend the atomic array by additional segments of atom arrays all in the ground state (see Fig.~\ref{fig1}), to the left and/or to the right. In the following, we will refer to the additional segments as atomic mirrors, and to the single excitation collective state in the central segment as a collective emitter. In order to follow the decay process, we will provide three probabilities. The first one is $p(t)$, the excitation probability shared in the whole atomic array the sample as defined by (\ref{2.2}) and (\ref{2.3}). The second one is $p_0(t)$ that captures the decay of the original excited state, as defined in Section~\ref{sec2B} by swapping $\hat{P}$ for $\hat{P}_0$. Finally we will also compute the probability $p_a(t)$:
\begin{equation}
p_a(t)=\mathrm{Tr}\, \hat{P}_a\hat{\rho}(t),%
\label{2.9}
\end{equation}
where $\hat{P}_a$ is defined as in (\ref{2.3}), with the only exception that it projects only onto the central segment subspace. In the considered one-dimensional open system, we can expect the recurrent incoherent scattering to play only a negligible role. This translates to the balance $p_0(t)=p_a(t)$ being fulfilled, which we will verify for any atomic configuration, within our calculation precision.

\section{Results in the Markovian regime} \label{sec3}

\noindent We now turn to studying the various configurations of the system with atomic Bragg mirrors added on the left and/or the right side. The response time of an $N$-atom Bragg mirror is $\Gamma_M^{-1}$, where $\Gamma_M \sim N\,\gamma_{\textrm{1D}}/2$ is similar to the cooperative decay rate introduced above. In this section we assume that the retardation over the length of the full array $L$ is limited compared to the response time of the atomic mirror, i.e., $L/v_g \ll \Gamma_M^{-1}$ where $v_g$ is the group velocity inside the waveguide. This is the so-called Markovian regime~\cite{Zheng2013,Shi2015,Scarani2016}, that is achieved when the distance $d_0$ between the segments is small. We first discuss (\emph{case 1}), where the full chain is in the atomic Bragg mirror configuration, i.e., $d_0=\lambda^{\mathrm{wg}}/2$. Then in (\emph{case 2}), $d_0=\lambda^{\mathrm{wg}}/4$ and we show that the breakdown of translational invariance for the full chain leads to massive changes in the response. Finally we comment on the temporal profiles and directionality of the light emitted into the waveguide in these different cases.

\subsection{Case 1: Bragg-mirror chain}

\begin{figure}[t!]
{${\includegraphics*[width=8.6cm]{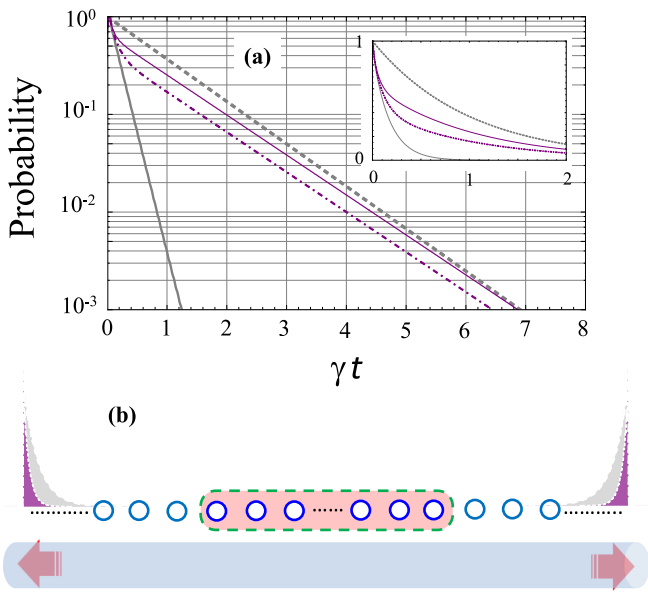}}$}
\caption{Translationally-invariant configuration with $d_0=\lambda^{\mathrm{wg}}/2$. The number of atoms are $N_L = N_C = N_R=100$. The asymmetric case with $N_L =0$ and $N_R = 200$ would yield the same results, as discussed in the text. (a) Decay of the probabilities $p_0(t)=p_a(t)$ (magenta dash-dotted) and $p(t)$ (magenta). Also shown are the cooperative emission of the unperturbed collective emitter $\exp(-\Gamma_C\,t)$ (gray) and the single-atom natural spontaneous decay in free space $\exp(-\gamma\,t)$ (gray dashed). The inset in linear scale clarifies the early stage of the decay process. (b) Schematic of the arrangement of the full chain and spatial profiles of the light pulses emitted into the waveguide (purple shaded profiles), see (\ref{2.7}). The grey shaded profiles represent the emission in the absence of atomic mirrors.
}
\label{fig2}%
\end{figure}%

\noindent In our first example we consider the configuration where $d_0=\lambda^{\mathrm{wg}}/2$. Here, the collective emitter consists in a segment of $N_C$ atoms from a longer array of $N_{\mathrm{tot}} = N_L + N_C + N_R$ atoms, that essentially forms an atomic Bragg mirror with $N_{\mathrm{tot}}$. Figure~\ref{fig2} shows numerical results for the decay probabilities, calculated with $N_L = N_C = N_R = 100$~atoms, i.e., for 300 atoms in total.

We first note that, in this situation, the decay dynamics is insensitive to the actual position of the collective emitter within the full array. For a small enough number of atoms such that the Markovian approximation also applies for propagation along the full array from end to end, the retardation for the internal virtual processes is negligible. The initially prepared superradiant mode can thus be instantly re-expanded in the basis set of the collective eigenmodes of the complete resolvent operator and then obey the wavepacket dynamics driven by its complete self-energy part (effective Hamiltonian).

The beginning of the decay reveals a stage of fast cooperative emission into the waveguide, which arises from the projection of the initial state onto the superradiant mode of the complete atomic system. This projection has amplitude $\sqrt{N_C/N_{\mathrm{tot}}}$, hence the initial fast decay in Fig.~\ref{fig2} lasts only until $p(t)$ has decreased by $N_C/N_{\mathrm{tot}}$.

The asymptotes of $p_0(t)$ and $p(t)$ deviate from the natural spontaneous decay $\exp(-\gamma\,t)$. These later time dynamics result from the projection onto the $(N_{\mathrm{tot}}-1)$ other collective modes, which are subradiant into the waveguide and hence decay by spontaneous emission into the external modes with a decay rate $\gamma^{\mathrm{ext}}$. The presence of the waveguide restricts spontaneous emission to a smaller than $4\pi$ solid angle, resulting in $\gamma^{\mathrm{ext}}<\gamma$. \footnote{If we combine the single-atom emission rate into the waveguide $\gamma_{\mathrm{1D}}$ with emission into the external modes $\gamma^{\mathrm{ext}}$ we get $\gamma^{\mathrm{ext}}+\gamma_{\mathrm{1D}}>\gamma$ (Purcell effect). Simultaneously we have $\gamma^{\mathrm{ext}}<\gamma$.} This is a specific property of this one-dimensional and translationally invariant atomic configuration, arising from the ability to cooperatively emit the photon along the array direction via the Rayleigh channel with only a negligible probability to scatter via the Raman channel. We found that $p_0(t)=p_a(t)$ within our numerical precision, showing that there is no signature of recurrent scattering in the emission process and that the atomic excitation in the central segment only occupies the superradiant mode state (\ref{2.8}).

\begin{figure}[t!]
{${\includegraphics*[width=8.6cm]{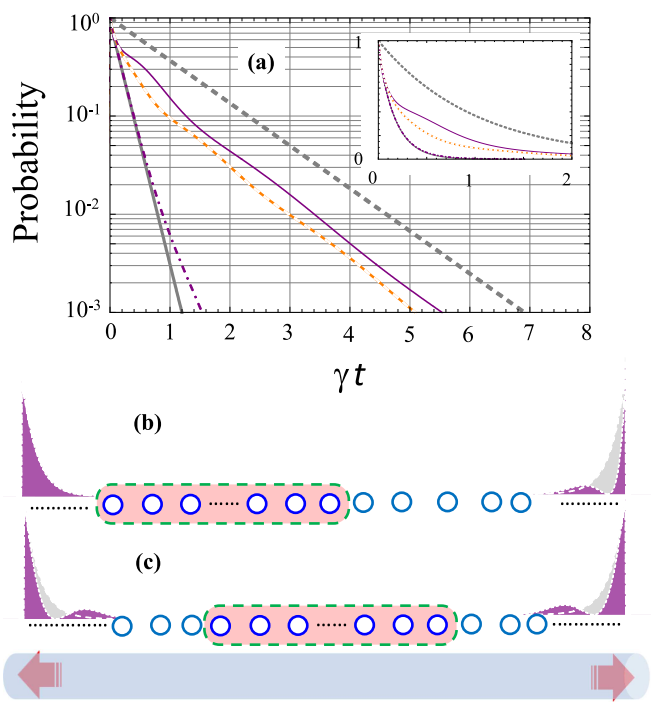}}$}
\caption{ Same as in Fig.~\ref{fig2} ($d_0\sim\lambda^{\mathrm{wg}}/2$, $N_{\mathrm{tot}} = 300$~total atoms) but with disordered atomic mirrors, and considering the cases of one-sided and symmetric arrangement of the atomic mirrors. (a) The decay probabilities $p_0(t)= p_a(t)$ (dash-dotted purple) coincide with the cooperative emission of the collective emitter alone. The excitation probability for the whole sample $p(t)$ decays more initially if the collective emitter is placed at one end of the full array [orange dotted, configuration shown in (b)] as opposed to between the two disordered atomic mirrors [purple, configuration shown in (c)]. (b) and (c) display the one-sided arrangement and the symmetric one, and provide the associated emission profiles.
}
\label{fig3}%
\end{figure}%

The perfect ordering of atoms is a crucial requirement for the data presented in Fig.~\ref{fig2}. Disorder in the left and right atomic mirrors, such as completely randomized positions, would break their ability for giving rise to coherent Bragg-type interference with the atoms in the central segment. In this case the decay dynamics is modified as shown in Fig.~\ref{fig3}, where the atoms in the mirrors are disordered, following a uniform probability distribution along the length of the mirror with a density of $1$ atom per $\lambda^{\mathrm{wg}}/2$. The initially excited atoms remain fully ordered. Somewhat surprisingly, the beginning stage of the decay process in Fig.~\ref{fig3} appears similar to the case of the ordered system shown in Fig.~\ref{fig2}. Even for the disordered configuration there is meaningful probability to project onto a cooperatively emitting collective state of the full atomic array involving all the atoms and with specific choice of the phase shifts $\vartheta_a$ in (\ref{2.8}). That collective state exhibits fast superradiant emission from the full array into the waveguide observed at short times in Fig.~\ref{fig3}.

The most striking difference with the ordered configuration concerns the population of the original state. In the tail of the process the dynamics changes and the disordered mirrors act as absorbers that incoherently re-scatter the photon into any mode. As a consequence we observe an exponential depopulation of the initial state nearly identical to the original undisturbed cooperative emission: the excitation leaves the collective emitter as if the atomic mirrors were absent. Just as above, no other collective state than the superradiant mode is populated within the active atoms, i.e. $p_0(t)= p_a(t)$.

The emission at long times, given by the asymptote of $p(t)$, is also different from the ordered case. In particular, a bump in the emission appears at long times in Fig.~\ref{fig3}(b), that has its counterpart in $p(t)$ in Fig.~\ref{fig3}(a), which originates mainly from the atomic mirrors as the excitation has depleted from the collective emitter ($p_a(t)$ is orders of magnitude smaller than $p(t)$ then). The long-term asymptote for $p(t)$ is an exponential decay $\sim \exp(-(\gamma^{\mathrm{ext}}+\gamma_{\mathrm{1D}})t)$, slightly faster than $\exp(-\gamma\,t)$, which originates from the Purcell enhancement of the single-atom emission by the presence of the waveguide. Interestingly, if the collective emitter is placed in the center of the full array rather than just on one end, the emission is delayed. Indeed if one end of the array is open for the emission from the initial state, the emitted light in that direction is directly lost and cannot be re-absorbed.

We presented our simulations for a particular random configuration of the atomic mirror, without ensemble averaging, so the emission of the sample indicates a slight signature of its configurational dependence. That is more visible for the approximately symmetric configuration showing slight beatings in its decay dynamics, see Fig.~\ref{fig3}(c).

\subsection{Case 2: Broken invariance}

\noindent Let us now consider a different configuration, where the translational symmetry along the full array is broken, but only at the boundary between the segments by shifting the separation $d_0$ from its reference value $\lambda^{\mathrm{wg}}/2$ to $\lambda^{\mathrm{wg}}/4$. At the same time we preserve the periodic distance of $\lambda^{\mathrm{wg}}/2$ between the atoms in each segment and their ability for the Bragg-type interference with the guided light. With identical left and right atomic mirrors, this configuation is then the extension of Refs.~\cite{Chang2012,Scarani2016} from a single atom to a collective emitter in the central segment of the chain.

In Refs.~\cite{Chang2012,Scarani2016} this configuration has been called a \emph{quantum cavity}, given that the central atoms sit at the anti-node of the photonic mode and therefore maximally couple to it. A mapping to the Jaynes-Cummings model was made~\cite{Chang2012}, exemplified by the appearance of a vacuum Rabi splitting, which shows that the density of photonic states is indeed enhanced. However we emphasize the crucial difference with cavity-QED, already pointed out in~\cite{Scarani2016}, that \emph{in this process the left and right segments neither operate as classical light reflectors nor as a cavity trapping the light}. This means that the photon lifetime in the system does not extend beyond the natural lifetime with the presence of the ``atomic Bragg mirrors''.

In this geometry we eliminate the contribution of the global superradiant mode, which still exists but only has negligible overlap with the initially prepared state. But now each atomic segment, considered independently, possesses a superradiant mode (\ref{2.8}) as one of the eigenstates of its own partial resolvent, and the complete resolvent operator can be decomposed into the basis of these states and reproduced by a separated matrix block in a part of the atomic Hilbert subspace. As a consequence the initial state, being a superposition of the eigenstates of this matrix block, would further evolve as a wave-packet, with periodic swapping of the excitation between the atomic segments. During this swapping process each segment has the ability for cooperative emission into the waveguide. The probability $p_0(t)$ should then reveal an oscillating behaviour, a signature of partial revival to the initial state. Physically it corresponds to a coupled dynamics between the atomic segments. During an excitation swapping cycle, a significant part of the optical excitation leaks into the waveguide. Since the emission into the waveguide has a superradiant nature from all segments, the total excitation probability $p(t)$ depletes faster than in the preceding examples, shown in Fig.~\ref{fig2} and Fig.~\ref{fig3}.

\begin{figure}[t!]
{${\includegraphics*[width=8.6cm]{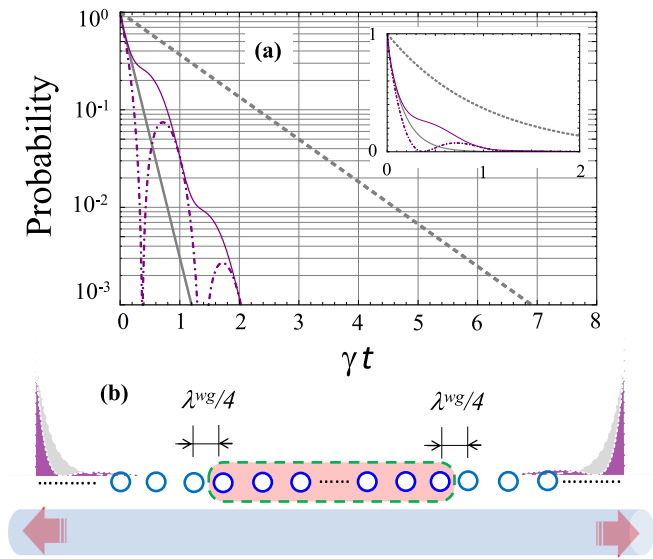}}$}
\caption{Same as in Fig.~\ref{fig2} but for $d_0=\lambda^{\mathrm{wg}}/4$. The atomic mirrors are added symmetrically on both sides.}
\label{fig4}%
\end{figure}%

\begin{figure}[tp]
{${\includegraphics*[width=8.6cm]{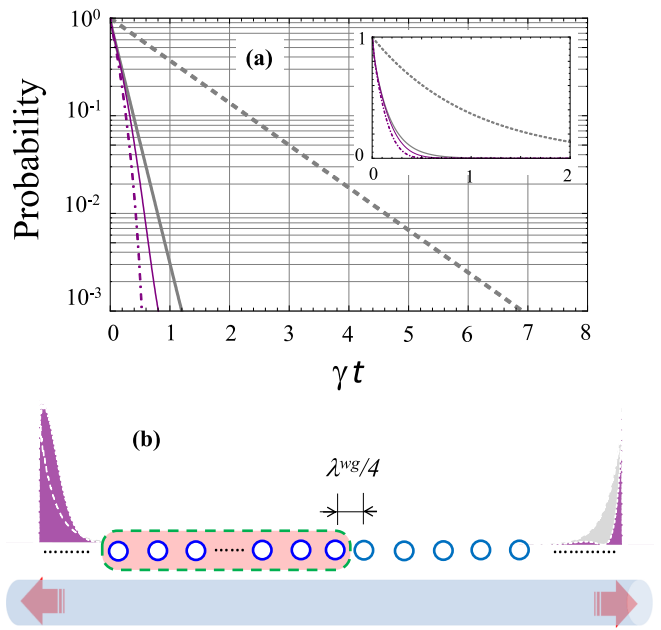}}$}
\caption{Same as in Fig.~\ref{fig4} but with one single atomic mirror with $N_R = 200$~atoms. Noticeably, both $p(t)$ and $p_0(t)$ decay faster than the bare collective emitter (grey solid line).
}
\label{fig5}%
\end{figure}

Figures~\ref{fig4} and \ref{fig5} present the result of our numerical simulations, supporting the above arguments for the same other calculation parameters as in Fig.~\ref{fig2}. The strong difference with the results of Fig.~\ref{fig2} is evident, with the above mentioned oscillating behavior being observable in Fig.~\ref{fig4}. These oscillations are the direct counterpart to the behavior observed in Refs.~\cite{Chang2012,Scarani2016} with a collective emitter instead of a single atom, here with full inclusion of dipole-dipole coupling and decay through free space. Noticeably the decay process is slowed down and its dynamics indicates the reviving oscillations for $p_0(t)= p_a(t)$ as well as certain deviations from exponential decay of $p(t)$.

When the atomic mirror is added to one side only, as shown in Fig.~\ref{fig5}, the excitation exits the system faster than from the bare collective emitter. In the beginning stage, approximately one half of the excitation is emitted directly into the waveguide towards the empty direction. The other part of the excitation transfers to the atomic mirror and is then cooperatively emitted into the waveguide at a higher rate. In such a geometry the reviving oscillations of $p_0(t)= p_a(t)$ only appear in a long-term asymptote beyond the scale of the graph reproduced in the plot.

\subsection{Emission into the waveguide}
\noindent Let us now briefly discuss the spatial profiles of the light emitted into the waveguide. For all of the considered configurations a significant portion of the emitted light is injected into the waveguide, since the initial state emits (at least partly) cooperatively inside the waveguide. In Figs~\ref{fig2}-\ref{fig5}, the purple shaded profiles show the light fractions propagating within the guided modes, while the grey-shaded profiles represent the emission into the waveguide by the collective emitter without atomic mirrors. Emission into both the azimuthal modes is included, but in fact the main part of the light wave-packet is transported in the $\sigma=-1$ azimuthal mode, which is strongly correlated with the radiation at $\sigma_{-}$ atomic transition, see Fig.~\ref{fig1}. Note that $p(t)$ amounts to these left and right emission profiles summed with the emission into free-space. For all configurations the addition of atomic mirrors atoms can either reduce or directionally enhance this emission. The differences in the outgoing pulse shape and in their energy balance between two propagating directions are quite sensitive to the location of the atomic mirrors with respect to the central segment. For instance in Fig.~\ref{fig3}(b), there is a strong imbalance of the cooperative emission into the waveguide towards the empty left side. This arises from the disordered chain acting as a light absorber. Figure~\ref{fig5} shows a similar directed emission, but its origin is fundamentally different. Here the excitation migrates between the segments and the directional dependence is associated with the cooperative emission into the waveguide from the two different segments independently, whereas in the case of Fig.~\ref{fig3}(b) the cooperative emission comes only from the collective emitter. The only exception to the dependence on the detailed configuration is the fully ordered, translationally symmetric atomic chain, shown in Fig.~\ref{fig2}, as discussed above.

In some cases, like with the geometry of Fig.~\ref{fig5}, the emission is found to decay faster than when the atomic mirrors are absent. This is a consequence of the collective effects from the full chain, which makes the cooperative emission into the waveguide generally stronger. The emission profiles thus further highlight the strong dependence of the system dynamics on the specifics of the atomic arrangement. Such directionality could for instance be useful for the implementation of efficient atom-based quantum optical protocols, as an alternative to intrinsic chiral systems~\cite{Lodahl2017,Jones2020}.

\begin{figure}[b!]
\includegraphics[width=8.6cm]{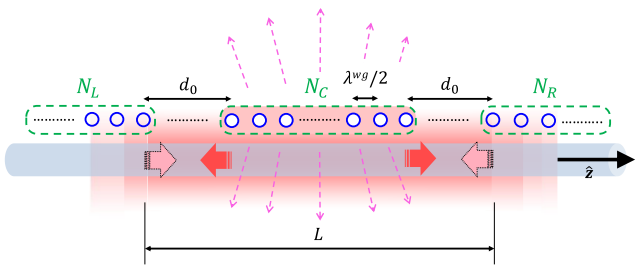}
\caption{An atomic Bragg cavity system of length $L$. In contrast to the previous studied cases where there was no photon lifetime enhancement, the left and right atomic mirrors are now separated by a \emph{long} distance $L$, as discussed in the text.}
\label{fig6}%
\end{figure}%

\section{Results in the non-Markovian regime}\label{sec4}

\noindent In the previous section, the behavior of the system could qualitatively be understood as the interaction of a collective emitter with atomic Bragg mirrors. In particular the symmetric situation of Fig.~\ref{fig4}, with $d_0= \lambda^{\mathrm{wg}}/4$, shares similarities with a classical cavity-QED configuration, with oscillations arising from a ``vacuum Rabi splitting''. Yet, the enhancement of photon lifetime, or photon trapping, a salient feature of optical resonators, does not arise for short atomic resonators.

The system can be transformed into a \emph{conventional} optical resonator if the left and right atomic mirrors are separated by a sufficiently long distance $L$, as shown on Fig.~\ref{fig6}. As commented in Appendix \ref{Appendix_Simplified}, such a cavity configuration requires that the time $L/v_g$ needed by a light pulse with a group velocity $v_g$ to cross the cavity is longer than the response time of the Bragg reflection from the atomic mirrors. The latter being $\Gamma_M^{-1}$, this condition translates to $L/v_g>\Gamma_M^{-1}$. This regime of waveguide-QED, where the retardation cannot be neglected, is commonly referred to as non-Markovian ~\cite{Zheng2013,Shi2015,Pichler2016,Sinha2020}.

Simultaneously, the coherence length of a photon emitted into the waveguide should be longer than the cavity length $L$ for the wave-packet to be trapped in the cavity volume and its field amplitude to be enhanced with respect to the open waveguide. With the coherence length given by $v_g/\Gamma_C^{-1}$ with $\Gamma_C^{-1}$ the duration of the emitted pulse in the open waveguide, this condition can be written as $L/v_g<\Gamma_C^{-1}$. These two inequalities ensure the closest physical conditions to the textbook model discussed in Appendix \ref{Appendix_Simplified} and they are simultaneously fulfilled if $\Gamma_M\gg \Gamma_C$, i.e., if the number of atoms  is much larger in the atomic mirrors than in the collective emitter.

In the following, we extend the study presented in Ref.~\cite{Scarani2016} by considering a collective emitter placed inside of a long cavity length $L$ made of atomic mirrors. We will consider the two distinct cases for $d_0$, as we did in Section~\ref{sec3} in the case of a short cavity. The first case is $d_0 = n \lambda^{\mathrm{wg}}/2$ (with $n$ being large positive integer number), where the atoms in the collective emitter are placed at nodes of a mode of the cavity formed by the atomic Bragg mirrors. By construction, this cavity mode is resonant with the atomic transition. The collective emitter is then partly decoupled from the waveguide because its constituent atoms are placed at nodes of the electric field of this resonant cavity mode. For the second case $d_0 = n \lambda^{\mathrm{wg}}/2 + \lambda^{\mathrm{wg}}/4$, the atoms in the collective emitter are placed at the anti-nodes of the electric field and therefore maximally couple to this cavity mode.

\begin{figure}[t!]
{${\includegraphics*[width=8.6cm]{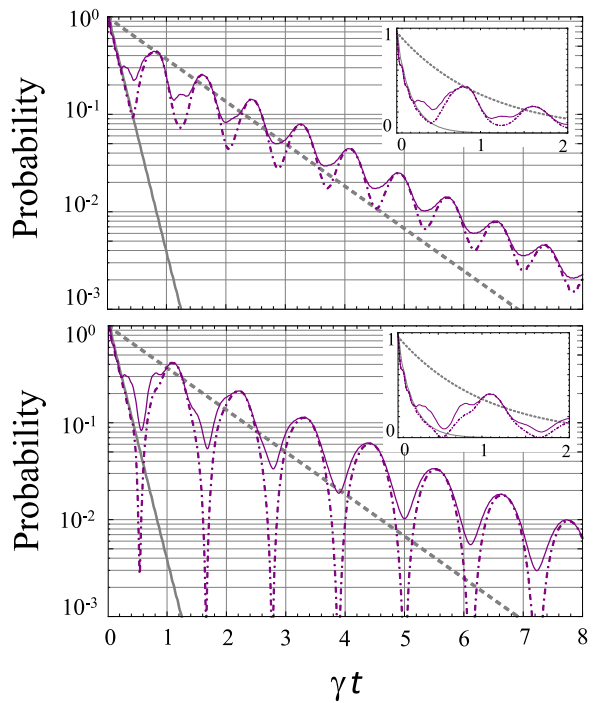}}$}
\caption{ Excitation probabilities in the case of a long resonator with $d_0\sim L/2=20$~cm and for $N_C = 100$~atoms and $N_L = N_R = 500$~atoms. Color and line type conventions are the same as in Figs.~\ref{fig2}-\ref{fig5}. (a) and (b) correspond to $d_0 = n \lambda^{\mathrm{wg}}/2$ and $d_0 = n\lambda^{\mathrm{wg}}/2 + \lambda^{\mathrm{wg}}/4$ (where $n$ is integer number) respectively.
}
\label{fig7}%
\end{figure}%

Figure~\ref{fig7} provides the results of our numerical simulations in these two cases, for $N_C = 100$~atoms, and $N_L = N_R = 500$~atoms for each atomic Bragg mirror. The atomic mirrors and collective emitter are separated by $d_0\sim L/2=20$~cm, approximately fulfilling the above requirements. In the weak coupling case shown in Fig.~\ref{fig7}(a) ($d_0 = n \lambda^{\mathrm{wg}}/2$), both $p(t)$ and $p_0(t)$ initially exhibit a fast decay at a rate of order $\Gamma_C$, which indicates the initially superradiant light emission through the distant chain segments. But the process is tailed asymptotically by a much slower and slightly oscillating decay. The observed weak oscillations appear as a result of $\Gamma_C\,L/v_g$ being finite, which imperfectly fulfills the condition on the emitted wave-packet, with the spectrum distributed within $\Gamma_C$. As a result, for sufficiently long cavity the atoms can still couple to the guided modes proximal to the resonant mode but without coupling to the resonant mode itself. In the strong coupling case shown in Fig.~\ref{fig7}(b) ($d_0 = n \lambda^{\mathrm{wg}}/2 + \lambda^{\mathrm{wg}}/4$), as expected in cavity-QED as well, the model predicts much stronger intra-cavity quantum Rabi oscillations.

The salient features of Rabi oscillations or decay at two time scales are therefore qualitatively similar for long and short cavities. There are however crucial differences with the behavior observed in Figs.~\ref{fig2}, \ref{fig4} and \ref{fig5}, related to the above-mentioned photon trapping, which justify the full cavity-QED behavior for long resonators. First of all, in Fig.~\ref{fig7} the probabilities $p_{0}(t)$ and $p(t)$ are nearly identical $p_{0}(t)\sim p(t)$. This indicates that the optical excitation is mainly in a polaritonic superposition of the superradiant atomic mode and intra-cavity field mode, and only a small part of the excitation is transferred into the atomic mirrors. As a consequence the incoherent losses from spontaneous emission into the external modes arise mostly from the collective emitter. In contrast with Section~\ref{sec3} there is no light leakage through the atomic mirrors into the open waveguide (i.e., emission from the atomic mirrors). It is also worth pointing out that the trapping of light within such an atomic Bragg cavity delays the emission into outer space~\cite{Scarani2016}, as demonstrated by the asymptotic exponential decay trends in Fig.~\ref{fig7}(b) being slower than in the previous examples of short resonators, as well as slower than single atom decay in external modes $\gamma^{\mathrm{ext}}$.

We have therefore shown that the regime of a classical cavity, exhibiting extended photon lifetimes, can also be achieved with a collective emitter provided that the cavity is long enough.
This was shown both in the regime of enhanced coupling through the cavity as well as in the regime of suppressing the emission into a selected guided mode.

\section{Conclusion}

\noindent In this paper we have theoretically studied the process of cooperative emission of a collective emitter into a nanoscale dielectric waveguide, in the presence of atomic Bragg mirrors coupled to the same waveguide. The atomic ensemble can be prepared and controlled with state-of-the-art experimental techniques. The approach is based on a three-dimensional quantum scattering theory adjusted to the specific geometry of a one-dimensional atomic array positioned along a waveguide, and primarily interacting with the evanescent field of the guided modes. The relatively simple atomic energy structure and the analytically-constructed, near-surface asymptote of the electric field Green's function allowed us to apply the rigorous three-dimensional analysis tracking all the main features of the internal coupling and configuration dependence of the emission process.

The numerical simulations show a strong sensitivity of the emission dynamics to the atomic arrangement. We have obtained that the emission process is expected to critically depend not only on the original localization of the optical excitation but also on the entire distribution of the atoms structuring the complete chain. The calculation results have predicted an oscillation behavior of the occupation probability and directional dependence of the pulse emitted into the waveguide. Our results with a collective emitter are consistent with a mapping to a Jaynes-Cummings Hamiltonian, as  previously described in the literature for a single emitter \cite{Chang2012,Scarani2016}.

In the specific long cavity setting, the system truly behaves as a cavity trapping the emitted light, where the edge atoms play the rule of \emph{classical} Bragg mirrors. This \emph{cavity regime} requires specific settings for the system parameters. We have pointed out the main criterion of it, namely the importance of low inertia in the reflection response of the Bragg-mirrors in comparison with the retardation associated with the virtual dynamics of the internal interactions. In this case we have observed quantum Rabi oscillations as well as slowing down of the external emission, which are convincing indicators that the light is trapped inside the cavity. Compared to previous works along these lines, these observations confirmed that these ideas extend well from fully 1D configurations with only one atom to a collective emitter and in the presence of dipole-dipole coupling and decay via free space.

Such systems have the potential for implementing tool of light trapping in one-dimensional cavity-type atomic structures, which would be of interest in the context of quantum information processing and interfacing. For example the coupling between atomic qubits stored in the spin degrees of freedom of atomic segments loaded inside such an atomic cavity would be significantly enhanced.

\acknowledgments
We are deeply grateful to Mark Havey for the long and fruitful collaborative work during many years. We thank him for many inspiring discussions of the problems considered in this paper.

This work was supported by the Russian Foundation for Basic Research under Grants No. 18-02-00265-A and No. 19-52-15001-CNRS-a, the Russian Science Foundation under Grant No. 18-72-10039, the Foundation for the Advancement of Theoretical Physics and Mathematics BASIS under Grant No. 18-1-1-48-1, the CNRS PRC France-Russia project Quantum1D, the French National Research Agency NanoStrong Project (ANR-18-CE47-0008), and the R\'egion Ile-de-France (DIM SIRTEQ). This project has also received funding from the European Union's Horizon 2020 research and innovation programme under grant agreement No. 899275 (DAALI project).

\appendix

\section{Computation of the dynamics}\label{Appendix_Dynamics}

\noindent The evolution operator $U(t,0)$ used in (\ref{2.1}) is given by the following inverse Fourier transform
\begin{equation}
U(t,0)=-\int_{-\infty}^{\infty}\frac{dE}{2\pi i}\frac{1}{E-\hat{H}+i0}\exp\left(-\frac{i}{\hbar}E\,t\right)%
\label{a.1}%
\end{equation}
where the integrand is the resolvent operator of the system Hamiltonian
\begin{equation}
\hat{R}(E)=\frac{1}{E-\hat{H}}%
\label{a.2}%
\end{equation}
acting in the full atoms-field Hilbert space of infinite dimension. The energy argument $E$ can be an arbitrary complex number parameter. It is however crucially helpful for the entire derivation that the resolvent operator actually contributes in (\ref{2.4}) only by being projected onto the finite subspace of the atomic subsystem
\begin{equation}
\tilde{\hat{R}}(E)=\hat{P}\,\hat{R}(E)\,\hat{P}.%
\label{a.3}%
\end{equation}
In order to evaluate the excitation probability, one thus needs to trace the operator product projected onto the subspace of large but finite dimension.

The resolvent operator (\ref{a.3}) can be evaluated with the invariant perturbation theory technique and with the Feynman diagram method. We refer the reader to the supplementary materials in~\cite{SMLK2012,KSH2017,SKGLK2015,PLGPCLK2018,Pivovarov2020} for a detailed presentation of the calculation approach. Here we only outline the most important calculation steps.

The resolvent (\ref{a.3}) can be expressed by a many-particle collective Green's function transporting the optical excitation through the atomic chain, see \cite{SMLK2012}. This many-particle Green's function fulfills a generalized Dyson-type equation presented in an integral form, whose kernel is expressed by the self-energy part, accumulating the complete internal field coupling between any atoms of the chain. In turn, each coupling diagram is given by the single-photon propagator, which near a nanoscale structure can be approximately built up in analytical form \cite{PLGPCLK2018}. Then, being considered for a collection of atoms at rrest, the many particle integral Dyson equation can be converted to the set of algebraic equations introduced directly for the reduced resolvent (\ref{a.3}). Finally the constructed system of algebraic equations can be numerically solved for a chain consisting of an arbitrary number of atoms.

\section{M{\o}ller  operator $\hat{\Omega}^{(-)\dagger}$}\label{Appendix_Moller}

\noindent Let us focus now on the operator $\hat{\Omega}^{(-)\dagger}$, which is the most important for our consideration since it directly contributes to the probability amplitude of the process. In accordance with its definition (\ref{2.5}), this operator can be expressed by the following time integral
\begin{equation}
\hat{\Omega}^{(-)\dagger}=-\frac{i}{\hbar}\int_0^{\tau/2}\!\!dt\,\exp\left(+\frac{i}{\hbar}\hat{H}_0\,t\right)\hat{V}\exp\left(-\frac{i}{\hbar}\hat{H}\,t\right)+\hat{I}%
\label{b.1}
\end{equation}
with $\tau\to +\infty$, and where $\hat{V}=\hat{H}-\hat{H}_0$ is the interaction Hamiltonian in the Schr\"{o}dinger time independent representation, and $\hat{I}$ is a unit operator.

The required matrix element taken between an arbitrary final state $|f\rangle$, consisting of the outgoing photon, and the initial state $|\psi(0)\rangle$, consisting of the excited atoms, is given by
\begin{eqnarray}
\lefteqn{\langle f|\hat{\Omega}^{(-)\dagger}|\psi(0)\rangle}%
\nonumber\\%
&&=-\frac{i}{\hbar}\int_0^{\tau/2}\!\!dt\,\langle f|\hat{V}\hat{P}\exp\left[\frac{i}{\hbar}\left(E_f-\hat{H}\right)t\right]\hat{P}|\psi(0)\rangle%
\nonumber\\%
\label{b.2}
\end{eqnarray}
where we have assumed orthogonality of the states $\langle f|\psi(0)\rangle=0$. Here, $E_f$ is the energy of the final state assumed as an eigenstate of $\hat{H}_0$. We have also multiplied and substituted $\hat{V}\to\hat{V}\hat{P}$, as made possible by the well-fulfilled assumption of the RWA approach.

The presence of the projector $\hat{P}$ in the integrand guarantees its depletion with time and provides rapid conversion of the integral itself. In turn, that lets us make further transfer up to the infinite upper limit $\tau/2\to +\infty$ as required by the definition of the M{\o}ller operator. We arrive to the following expression for the matrix element
\begin{equation}
\langle f|\hat{\Omega}^{(-)\dagger}|\psi(0)\rangle=\langle f|\hat{V}\tilde{\hat{R}}(E_f+i0)|\psi(0)\rangle%
\label{b.3}%
\end{equation}
where the contributing resolvent operator is defined by (\ref{a.3}) and (\ref{a.2}).

\section{Interaction Hamiltonian}\label{Appendix_Interaction}
\noindent The atoms-field interaction near a nanostructure is based on a general paradigm of statistical physics \cite{LfPtIX}, which states that the photon propagator can be expressed by the retarded-type fundamental solution of the macroscopic Maxwell equation
\begin{eqnarray}
\lefteqn{\triangle D^{(R)}_{\mu\nu}(\mathbf{r},\mathbf{r}';\omega) -\frac{\partial^2}{\partial x_{\mu}\partial x_{\alpha}}D^{(R)}_{\alpha\nu}(\mathbf{r},\mathbf{r}';\omega)}
\nonumber\\%
&+&\frac{\omega^2}{c^2}\left[1+4\pi\chi(\mathbf{r})\right]D^{(R)}_{\mu\nu}(\mathbf{r},\mathbf{r}';\omega)=4\pi\hbar\,\delta_{\mu\nu}\delta(\mathbf{r}-\mathbf{r}')\,.%
\nonumber\\%
\label{c.1}
\end{eqnarray}
Here $\chi(\mathbf{r})$ is the dielectric susceptibility of the medium (extended by $\chi(\mathbf{r})=0$ in free space). We use the Cartesian coordinates $\mu,\nu,\alpha$ and assume the sum over the repeated indices. In the dipole gauge, the quantum description of the interaction process implies a virtual exchange by the excitation between atoms provided by the causal-type Green's function of the electric field. If this function is known, the resolvent operator (\ref{a.3}) can be compiled by a perturbation theory expansion based on the second-quantization formalism, see \cite{SMLK2012,KSH2017,SKGLK2015,PLGPCLK2018}.

For a nanoscale waveguide as considered here, the causal-type Green's function for the evanescent part of the electric field (distributed outside the nanofiber) is assumed to be linked with the photon propagator by the following transformation
\begin{eqnarray}
D^{(E)}_{\mu\nu}(\mathbf{r},\mathbf{r}';\omega)\!\!&=&\!\!-i\!\int^{\infty}_{-\infty}\!\!d\tau\,%
\mathrm{e}^{i\omega\tau}\!\left.\langle T E_{\mu}(\mathbf{r},t)\,E_{\nu}(\mathbf{r}',t')\rangle\right|_{\tau=t-t'}%
\nonumber\\%
\!&=&\!\frac{\omega^2}{c^2}D^{(R)}_{\mu\nu}(\mathbf{r},\mathbf{r}';|\omega|)%
\label{c.2}
\end{eqnarray}
where we have assumed a lossless transparent dielectric medium with a dielectric permittivity $\epsilon(\mathbf{r})=1+4\pi\chi(\mathbf{r})$.

As shown in \cite{PLGPCLK2018}, the above equations can be fulfilled once we re-expand the electric field operator, originally introduced in the Schr\"{o}dinger representation, in the basis of the guided and external modes
\begin{equation}
\hat{\mathbf{E}}(\mathbf{r})=\sum_{s}\left(2\pi\hbar\omega_s\right)^{1/2}\!%
i\left[b_s\mathbf{E}^{(s)}(\mathbf{r})%
- d_s^{\dagger}\mathbf{D}^{(s)\ast}(\mathbf{r})\right]\ + \ldots%
\label{c.3}%
\end{equation}%
Here we have selected only the contribution of the guided modes, specified by the mode index $s$ and frequency $\omega_s$, for the electric field $\mathbf{E}^{(s)}(\mathbf{r})$ and the displacement field $\mathbf{D}^{(s)}(\mathbf{r})=\epsilon(\mathbf{r})\,\mathbf{E}^{(s)}(\mathbf{r})$, where $\epsilon(\mathbf{r})$ is the spatially dependent dielectric permittivity of the entire medium (free space and dielectric waveguide).  The ellipsis in Eq.~(\ref{c.3}) denote the contribution of the external modes. The mode functions are normalized as
\begin{eqnarray}
\lefteqn{\int\! d^3r\,\epsilon(\mathbf{r})\, \mathbf{E}^{(s')\ast}(\mathbf{r})\cdot\mathbf{E}^{(s)}(\mathbf{r})}
\nonumber\\%
&&\equiv\int\! d^3r\,\mathbf{D}^{(s')\ast}(\mathbf{r})\cdot\mathbf{E}^{(s)}(\mathbf{r})=\delta_{s's}%
\label{c.4}
\end{eqnarray}
The displacement field has a transverse profile verifying $\mathrm{div}\mathbf{D}^{(s)}(\mathbf{r})=0$, which is an important requirement for the transformation (\ref{c.3}). The mode operators $b_s$ and $d^\dagger_s$ approximately obey the standard bosonic commutation rules $[b_s,d_{s'}^{\dagger}]=\delta_{s,s'}$ \cite{PLGPCLK2018}. This furthermore implies the standard approach of the invariant perturbation theory with applicability of the Wick theorem and diagram expansion. The supporting arguments that the microscopic quantum description stays applicable in such a slightly modified form follow from our basic assumptions. Physically, it is a consequence of the fact that the sub-wavelength nanoscale object can only weakly affect the original vacuum structure of the field, such that the field in any waveguide mode mostly exists in its outer evanescent part.

The interaction of the quantized field with the chain is taken in the long wavelength dipole approximation
\begin{eqnarray}
\hat{V}&=&-\sum_{a=1}^{N}%
\hat{\mathbf{d}}^{(a)}\cdot\hat{\mathbf{E}}(\mathbf{r}_a)+\hat{H}_{\mathrm{self}}.%
\label{c.5}%
\end{eqnarray}
The first term accumulates partial interactions for each of an $a$-th atomic dipole $\mathbf{d}^{(a)}$ with the electric field $\hat{\mathbf{E}}(\mathbf{r})$ (\ref{b.3}) at the point of the dipole position. The second term is the dipoles' self-energy part, which is mainly important for renormalization of the self-action divergencies \cite{CohTann92,SKKH2009,KSH2017}. In the dipole gauge, the operator (\ref{c.3}) represents the microscopic displacement field, which coincides with the electric field at a certain distance from the dipole. The difference, as well as the contribution of the second term in (\ref{c.5}), would be important for evaluation of the single particle self-energy associated with the interaction of the dipole with its own field near a nanostructure.

\section{Simplified model of the decay process}\label{Appendix_Simplified}
\noindent Here we compare our results with an analytically solvable model of a single atom interacting with a cavity mode \cite{Scully97}. The atom, storing a single excitation, will model the collective emitter discussed in the main text. The atom is described by the following set of Pauli dyad-type operators
\begin{eqnarray}
\sigma_{z}&=&|e\rangle\langle e| - |g\rangle\langle g|%
\nonumber\\%
\sigma_{+}&=&|e\rangle\langle g|%
\nonumber\\%
\sigma_{-}&=&|g\rangle\langle e|%
\label{d.1}%
\end{eqnarray}
acting in a two-dimensional energy subspace of the atomic ground $|g\rangle$ and excited $|e\rangle$ states. Let the atom be coupled with a single cavity mode expressed by the photon annihilation and creation operators $c$ and $c^{\dagger}$ respectively. In turn, the light in the cavity mode can leak through the cavity mirrors and transform into the incoming and outgoing field waves associated with the mode propagating in free space and expressed by operators $a$ and $a^{\dagger}$.

Then, as shown in textbooks \cite{Gardiner91,Scully97}, in the Heisenberg picture the system obeys the following set of the Heisenberg-Langevin equations
\begin{eqnarray}
\dot{\sigma}_{z}&=&-2g_a\left[c^{\dagger}\sigma_{-}+\sigma_{+}c\right]%
\nonumber\\%
\dot{\sigma}_{-}&=&+g_a\,\sigma_z\,c%
\nonumber\\%
\dot{\sigma}_{+}&=&+g_a\,c^{\dagger}\,\sigma_z%
\nonumber\\%
\dot{c}&=&-\frac{\kappa}{2}\, c + g_a\sigma_{-}-\sqrt{\kappa}\,a_0(t)%
\nonumber\\%
\dot{c}^{\dagger}&=&-\frac{\kappa}{2}\, c^{\dagger} + g_a\sigma_{+}-\sqrt{\kappa}\,a_0^{\dagger}(t)%
\label{d.2}%
\end{eqnarray}
where $g_a$ is a coupling constant between the atom and cavity mode and $\kappa$ is a leakage rate through the cavity mirrors. All the operators are time dependent. For the Langevin noise terms in the last two equations, we emphasize that the field operators outside the cavity are considered in the interaction representation by specifically indicating the time-dependence, as well as adding a $0$-subscript.

The equations (\ref{d.2}) can be solved for the first statistical moments and lead to the following dynamics of the upper state population probability $p=p(t)$
\begin{eqnarray}
\lefteqn{p(t)=\frac{1+\langle\sigma_z(t)\rangle}{2}}%
\nonumber\\%
&&=\frac{2\exp\left(-\kappa t/2\right)}{\kappa^2-16g_a^2}\left\{-4g_a^2+\left[\frac{\kappa^2}{4}-2g_a^2+\frac{\kappa}{4}\left(\kappa^2-16g_a^2\right)^{1/2}\right]\right.
\nonumber\\%
&&\times\exp\left[\left(\kappa^2-16g_a^2\right)^{1/2}\frac{t}{2}\right]
+\left[\frac{\kappa^2}{4}-2g_a^2-\frac{\kappa}{4}\left(\kappa^2-16g_a^2\right)^{1/2}\right]%
\nonumber\\%
&&\left.\times\exp\left[-\left(\kappa^2-16g_a^2\right)^{1/2}\frac{t}{2}\right]\right\}\,.%
\label{d.3}%
\end{eqnarray}
In the two opposite limits of either low or high quality cavities, this respectively results to either enhanced spontaneous decay, or in Rabi-type oscillations initiated by the excitation swapping between the cavity field and atom. The introduced model can be generalized if instead of a single atom, we assume a collective atomic emitter consisting of $N$ atoms with the renormalized coupling constant $g_a\to \sqrt{N} g_a=g_{C}$ in solution (\ref{d.3}).

Let us make some useful estimates based on the scaling of this model with the parameters of the atomic chain and of the subwavelength waveguide, which we have discussed in the main text. We suppose the cavity as constructed by such a waveguide and two atomic Bragg mirrors separated by a distance $L$. Then, we can estimate the rate of the cavity losses as
\begin{equation}
\kappa\sim (1-R)v_g/L
\label{d.4}%
\end{equation}
where $v_g$ is the group velocity of the guided wave, and $R$ is the reflection coefficient of the mirrors, which is expected to be close to 100\%. For the considered atomic Bragg mirrors the strong reflection can be actually provided only in a quite narrow spectral domain near the point of atomic resonance $\omega_0$ where we approximately get
\begin{equation}
R(\omega)\sim\left|\frac{-\displaystyle\frac{i}{2}\Gamma_{M}}{(\omega-\omega_0)+\displaystyle\frac{i}{2}\Gamma_{M}}\right|^{2}%
\label{d.5}%
\end{equation}
where $\Gamma_{M}$ is the cooperative decay rate for the atomic array forming the mirrors. If the collective emitter were loaded between such mirrors separated by a short distance $L<v_g\,\Gamma_{M}^{-1}$, then the virtual interaction processes would expand so fast that the entire system, consisting of the full chain, would dynamically transform to the collective quantum state before it would lose its excitation and the photon would be emitted into the outer space. Such a configuration cannot be relevantly described by the above model and can be only numerically simulated the way it is discussed in the main text.

If we assume the alternative configuration where the separation between the mirrors is sufficiently large such that $L>v_g\,\Gamma_{M}^{-1}$, then in accordance with (\ref{d.4}) we get
\begin{equation}
\kappa< (1-R)\Gamma_{M}\,.%
\label{d.6}%
\end{equation}
The cooperative decay of the collective emitter into the open cavity $\Gamma_C$ is faster than the benchmark given by (\ref{d.6}), following (\ref{d.5}), at least within a representative spectral domain of $\omega-\omega_0\sim\Gamma_C$. Thus we can expect the behavior of the compound system to be approximated by the solution (\ref{d.3}) taken above the critical point with $g_{C} > \kappa/4$. That predicts a signature of the intra-cavity Rabi-oscillations, as we have observed in our numerical simulations.

The important consequence of the above result is that the trend of the dynamics will not be affected by the absolute value of the coupling constant $g_C$ as long as the inequality holds. This means that the observation of Rabi oscillations will not depend on the external parameters such as the distance of the atoms from the surface, or the number of atoms in the active chain. However, as soon as the reverse inequality $g_C<\kappa/4$ is fulfilled, it critically change the structure of the solution (\ref{d.3}), and the decay rate has the same dependence on $g_C$ as the open system. For the parameters used in our paper, which we try to keep close to experimental conditions, the critical point $g_C\sim\kappa/4$ would be difficult to reach since (\ref{d.3}) and (\ref{d.6}) leads to cavity losses comparable with or even lower than the natural decay for a single atomic emitter. In this limit, reducing the number of active atoms down to a few atoms would lead to the loss of the main advantage of cooperative enhancement, which is crucially required to provide the preferable coupling with the guided mode.

\bibliography{references}

\end{document}